\documentclass{aa}

\usepackage{txfonts}
\usepackage{textcomp}

\usepackage{graphicx}	
\usepackage{amsmath}	
\usepackage{amssymb}	
\usepackage[alsoload=hep,separate-uncertainty=true]{siunitx}

\usepackage[caption=false]{subfig}
\usepackage{booktabs}

\usepackage[normalem]{ulem}

\usepackage{xcolor}

\usepackage{footmisc}

\usepackage{natbib}
\usepackage{hyperref}
\usepackage{cleveref}

\graphicspath{
    {plots/}
}

\definecolor{alizarin}{rgb}{0.82, 0.1, 0.26}
\definecolor{blue(munsell)}{rgb}{0.0, 0.5, 0.69}
\definecolor{asparagus}{rgb}{0.53, 0.66, 0.42}

\newcommand{\alphaFe}[0]{[$\alpha$/Fe]}
\newcommand{\steckmap}[0]{\texttt{STECMAP}}
\newcommand{\ppxf}[0]{\texttt{pPXF}}
\newcommand{\Ha}[0]{H$\alpha$}

\DeclareSIUnit\Gyr{Gyr}
\DeclareSIUnit\Myr{Myr}
\DeclareSIUnit\Mpc{Mpc}
\DeclareSIUnit\kpc{kpc}
\DeclareSIUnit\pc{pc}
\DeclareSIUnit\msol{\text{\ensuremath{\textup{M}_{\odot}}}}
\DeclareSIUnit\riso{\text{\ensuremath{\textup{R}_{25.5}}}}
\DeclareSIUnit\Rnb{\text{\ensuremath{\textup{R}_{\textup{nb}}}}}
\DeclareSIUnit\dex{dex}
\DeclareSIUnit\mag{mag}
\DeclareSIUnit\AB{AB}
\DeclareSIUnit\erg{erg}
\DeclareSIUnit\arcsec{arcsec}
\DeclareSIUnit\arcmin{arcmin}

\begin{document}

    \title{Galaxies within galaxies in the TIMER survey: stellar populations of inner bars are scaled replicas of main bars}
    \titlerunning{Stellar populations in inner bars}

    \author{%
        Adrian Bittner
        \inst{1,2}\and
        Adriana de Lorenzo-Cáceres
        \inst{3,4}\and
        Dimitri A. Gadotti
        \inst{1}\and
        Patricia Sánchez-Blázquez
        \inst{5,6}\and
        Justus Neumann
        \inst{7}\and
        Paula Coelho
        \inst{8}\and
        Jesús Falcón-Barroso
        \inst{3,4}\and
        Francesca Fragkoudi
        \inst{9}\and
        Taehyun Kim
        \inst{10}\and
        Ignacio Martín-Navarro
        \inst{3,4}\and
        Jairo Méndez-Abreu
        \inst{3,4}\and
        Isabel Pérez
        \inst{11,12}\and
        Miguel Querejeta
        \inst{13}\and
        Glenn van de Ven 
        \inst{14}
    }
    \authorrunning{A. Bittner et al.}

    \institute{%
        European Southern Observatory, Karl-Schwarzschild-Str. 2, D-85748 Garching bei München, Germany\\ \email{adrian.bittner@eso.org}\and
        Ludwig-Maximilians-Universität, Professor-Huber-Platz 2, 80539 Munich, Germany\and
        Instituto de Astrofísica de Canarias, Calle Vía Láctea s/n, E-38205 La Laguna, Tenerife, Spain\and
        Departamento de Astrofísica, Universidad de La Laguna, E-38200 La Laguna, Tenerife, Spain\and
        Departamento de Física de la Tierra y Astrofísica, Universidad Complutense de Madrid, E-28040 Madrid, Spain\and
        Instituto de Física de Partículas y del Cosmos, Universidad Complutense de Madrid, E-28040 Madrid, Spain\and
        Institute of Cosmology and Gravitation, University of Portsmouth, Burnaby Road, Portsmouth PO1 3FX, United Kingdom\and
        Instituto de Astronomia, Geofísica e Ciências Atmosféricas, Universidade de São Paulo, \\ R. do Matão 1226, 05508-090 São Paulo, Brazil\and
        Max-Planck-Institut für Astrophysik, Karl-Schwarzschild-Str. 1, D-85748 Garching bei München, Germany\and
        Department of Astronomy and Atmospheric Sciences, Kyungpook National University, Daegu 702-701, Korea\and
        Departamento de Física Teórica y del Cosmos, Universidad de Granada, Facultad de Ciencias, E-18071, Granada, Spain\and
        Instituto Universitario Carlos I de Física Teórica y Computacional, Universidad de Granada, E-18071 Granada, Spain\and
        Observatorio Astronómico Nacional, C/Alfonso XII 3, Madrid E-28014, Spain\and
        Department of Astrophysics, University of Vienna, Türkenschanzstraße 17, 1180 Wien, Austria
    }

    \date{Received; accepted}

\abstract{%
Inner bars are frequent structures in the local Universe and thought to substantially influence the nuclear regions of
disc galaxies. 
%
In this study we explore the structure and dynamics of inner bars by deriving maps and radial profiles of their mean
stellar population content and comparing them to previous findings in the context of main bars. To this end, we exploit
observations obtained with the integral-field spectrograph MUSE of three double-barred galaxies in the TIMER sample. 
%
The results indicate that inner bars can be clearly distinguished based on their stellar population properties alone.
More precisely, inner bars show significantly elevated metallicities and depleted {\alphaFe} abundances.  Although they
exhibit slightly younger stellar ages compared to the nuclear disc, the typical age differences are small, except at
their outer ends.  These ends of the inner bars are clearly younger compared to their inner parts, an effect known from
main bars as orbital age separation. In particular, the youngest stars (i.e.\ those with the lowest radial velocity
dispersion) seem to occupy the most elongated orbits along the (inner) bar major axis.  We speculate that these distinct
ends of bars could be connected to the morphological feature of ansae.  Radial profiles of metallicity and {\alphaFe}
enhancements are flat along the inner bar major axis, but show significantly steeper slopes along the minor axis. This
radial mixing in the inner bar is also known from main bars and indicates that inner bars significantly affect the
radial distribution of stars. 
%
In summary, based on maps and radial profiles of the mean stellar population content and in line with previous TIMER
results, inner bars appear to be scaled down versions of the main bars seen in galaxies. This suggests the picture of a
``galaxy within a galaxy'', with inner bars in nuclear discs being dynamically equivalent to main bars in main galaxy
discs.  
}
%
%
%


    \maketitle


\section{Introduction}%
\label{sec:introcution}
Bars are a frequent structure in disc galaxies and important for their secular evolution \citep[see
e.g.][]{kormendy2004, athanassoula2005, barazza2008, sheth2008, aguerri2009, masters2011, kraljic2012, fragkoudi2020}. But some
galaxies indeed host more than only one bar: such double-barred systems have a large-scale bar located in their main
discs, while another, smaller bar (which we will refer to as \textsl{inner bar}\footnote{In the literature inner bars
are often also referred to as \textsl{nuclear bars}. We warn the reader that sometimes the term nuclear bar is also used
to refer to particularly small bars in single-barred systems.} throughout this paper) can be found in their centres. The
first double-barred galaxies were already discovered in the 1970s \citep[NGC\,1291; ][]{deVaucouleurs1974,
deVaucouleurs1975} and early on these objects were thought to be part of a small group of dynamically peculiar galaxies.
However, some studies suggest that 30\% of all barred galaxies are actually double-barred systems \citep[see
e.g.][]{erwin2002, laine2002, erwin2004, buta2015}, while more recently Hildebrandt et al. (subm.) 
estimate a lower limit for the volume-corrected fraction of double bars in the CALIFA survey of \SI{12}{\percent}.

Despite the amount of known double-barred galaxies and the important secular processes inner bars might induce
\citep{shlosman1989, shlosman1990}, to date, few studies have explored their stellar population content in detail.
The first attempt was made by \citet{deLorenzoCaceres2012} using long-slit observations of the galaxy NGC\,357. Their
results indicate that the bulge and inner bar have similar stellar population properties while the main bar is less
metal-rich and more {\alphaFe} enhanced. Subsequent integral-field spectroscopic observations
\citep{deLorenzoCaceres2013} of four additional galaxies are compatible with these results, but further show that inner
bars appear slightly younger than their surroundings. Radial profiles reveal positive age and negative metallicity
gradients along both the inner and main bars, while {\alphaFe} abundances are flat. 

The TIMER project \citep[Time Inference with MUSE in Extragalactic Rings;][]{gadotti2019} is a survey focussing on
central structures, for instance nuclear discs, nuclear rings, and inner bars, observed in massive, barred disc galaxies
in the local Universe.  To date, 21 galaxies of the sample have been observed with the Multi-Unit Spectroscopic Explorer
\citep[MUSE; ][]{bacon2010} at the Very Large Telescope. While the main goal of the project is to infer the epoch of bar
formation from the star formation histories in the central components, previous TIMER studies also explored the nature
of inner bars in greater detail.  \citet{mendezAbreu2019} investigated the face-on galaxy NGC\,1291 and, for the first
time, detected kinematic signatures of a box/peanut structure associated with the inner bar in this galaxy. More
precisely, bi-symmetric minima of the higher-order moment $h_4$ of the line-of-sight velocity distribution are observed
along the inner bar major axis, as expected from numerical simulations and identified in the context of main bars
\citep[see e.g.][]{debattista2005, mendezAbreu2008, mendezAbreu2014}.  Similarly, \citet{bittner2019} observed a
correlation between radial velocity and the higher-order moment $h_3$ in the spatial region of the inner bar of
NGC\,1433. This correlation is a well-known kinematic signature of main bars, arising from the strongly elongated $x_1$
orbits within them \citep[][]{bureau2005, iannuzzi2015, li2018, gadotti2020}. Detecting this correlation for an
inner bar again suggests that both types of bars are dynamically similar structures. Finally,
\citet{deLorenzoCaceres2019} combined multi-component photometric decompositions of NGC\,1291 and NGC\,5850 with
measurements of kinematics and the stellar population content. The results show that nuclear discs and inner bars have
similar radii, suggesting that inner bars form via dynamical instabilities in nuclear discs in the same way main bars
arise in the main discs of galaxies.  In addition, the star formation histories indicate that the inner bars in
NGC\,5850 and NGC\,1291 are at least \SI{4.5}{\Gyr} and \SI{6.5}{\Gyr} old, implying that inner bars are dynamically
stable, long-lived structures. 

The stellar population properties of main bars have been studied in greater detail. \citet{perez2007} and
\citet{perez2009} detected large variations of ages and metallicities in bars, with positive, null, and negative
metallicity gradients along their major axis. \citet{sanchezBlazquez2011} compared the stellar content of bars with that
in the main discs of these galaxies, indicating that bars host, on average, older stellar populations with higher
metallicities than the galaxy discs. The gradients in both parameters appear flatter along the bar major axis compared
to the gradients found in the main disc. Using 128 galaxies from the MaNGA survey, \citet{fraser-mcKelvie2019} found
that the gradients of age and metallicity along the bar are typically flatter compared to the region outside of the bar
but within the bar radius \citep[see also][]{seidel2016}. \citet{neumann2020} exploited the high spatial resolution
observations of the TIMER survey to explore stellar population gradients not only along the bar major axis, but also
perpendicular to it. The results show that the youngest stellar populations in bars are located closest to the bar major
axis, suggesting that the youngest stars populate the most elongated orbits. 

The large number of morphological features in disc galaxies, especially in their centres, has led astronomers to
introduce a variety of nomenclatures for these structures.  In this study, we use the term \textsl{nuclear disc} to
refer to rotationally supported central discs with typical sizes of a few hundred parsecs that are kinematically
distinct from the main galaxy disc.  We choose this nomenclature to distinguish nuclear discs from inner and outer
discs, as often referred to in the context of disc breaks. The outermost edge of nuclear discs are often distinguished
by gaseous and/or star-forming rings. These structures are named \textsl{nuclear rings}, in order to discriminate them
from inner rings typically found at the radius of the main bar and outer rings located well outside of the main bar. For
the double-barred systems presented here, we use the term \textsl{inner bar} when referring to small-scale bars
associated with nuclear discs while large-scale bars found in the main discs of galaxies are simply denoted \textsl{bar}
or \textsl{main bar}, as appropriate. 

In this paper, we investigate the similarities between the stellar populations of main bars and inner bars by exploiting
two-dimensional stellar population maps and radial profiles, based on the high-resolution, integral-field spectroscopic
observations of the TIMER survey. We further complement the study of \citet{deLorenzoCaceres2019} by exploring
how the stellar population properties vary along and perpendicular to inner bars. In particular, we present for the
first time radial profiles of stellar population properties along the inner bar major and minor axis. We then compare
these results to those obtained by \citet{neumann2020} in the context of main bars. Based on these investigations, we
explore the dynamical structure of inner bars, argue that main and inner bars appear to be scaled versions of each
other, and speculate about the nature of ansae. 

This paper is structured as follows: in the next section we summarise observations, data reduction, sample selection,
and analysis which are already detailed in previous papers. In Sect.~\ref{sec:results} we show our results on the
stellar population content of inner bars before discussing them in the context of main bars in
Sect.~\ref{sec:discussion}. Finally, we summarise our conclusions in Sect.~\ref{sec:summary}. 

%
%
%

\section{Sample, observations, and data analysis}%
\label{sec:data}
In this section we briefly summarise our sample of inner bars, the TIMER observations and data reduction, as well as the
performed data analysis. 

%
%
%
\subsection{Double-barred galaxies in TIMER}%
\label{subsec:sample}
The TIMER sample is selected from the \emph{Spitzer} Survey of Stellar Structures in Galaxies \citep[S$^4$G,
][]{sheth2010}.  Therefore, all galaxies are nearby ($d < \SI{40}{\Mpc}$), bright ($m_B < \SI{15.5}{\mag}$), and large
($D_{25} > \SI{1}{\arcmin}$) objects.  In addition, only barred galaxies with central structures, as determined by
\citet{buta2015}, stellar masses above \SI{e10}{\msol}, and inclinations below \SI{60}{\deg} are included. 

Among the 21 TIMER galaxies observed so far, only NGC\,1291 and NGC\,5850 exhibit prominent inner bars. These inner bars
have already been detected in previous studies on the subject \citep[see e.g.][]{erwin2004, buta2015,
deLorenzoCaceres2008, deLorenzoCaceres2013} and recently investigated within the TIMER survey
\citep{deLorenzoCaceres2019, mendezAbreu2019}. 
In contrast, the presence of an inner bar in the centre of NGC\,1433
remains debated: previous photometric studies \citep{buta1986, jungwiert1997, erwin2004, buta2015} show some evidence of
an inner bar and \citet{bittner2019} finds consistent kinematic signatures. However,
\citet{deLorenzoCaceres2019} inspected recent \emph{Hubble} Space Telescope images of the galaxy and concluded that
NGC\,1433 does not host an inner bar. They argue that the presence of a variety of central structures, such as a nuclear
ring, nuclear disc, and nuclear spiral arms has led to a misclassification in this galaxy. Moreover, they state that the
absence of clear $\sigma$-hollows, a kinematic feature typically associated with inner bars, further suggests that this
galaxy does not host an inner bar. 

\begin{figure*}
    \centering
    \includegraphics[width=0.33\textwidth]{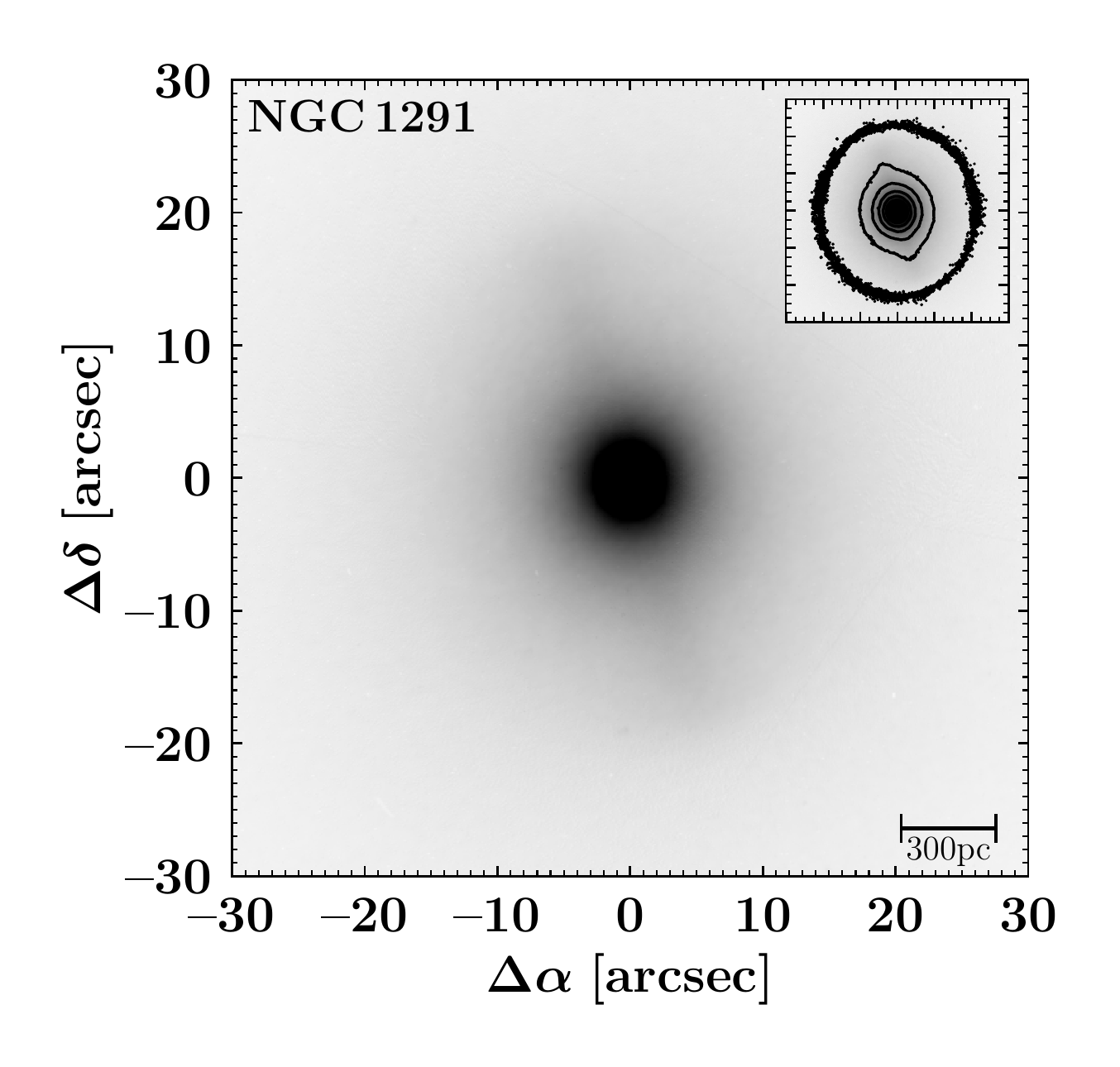}
    \includegraphics[width=0.33\textwidth]{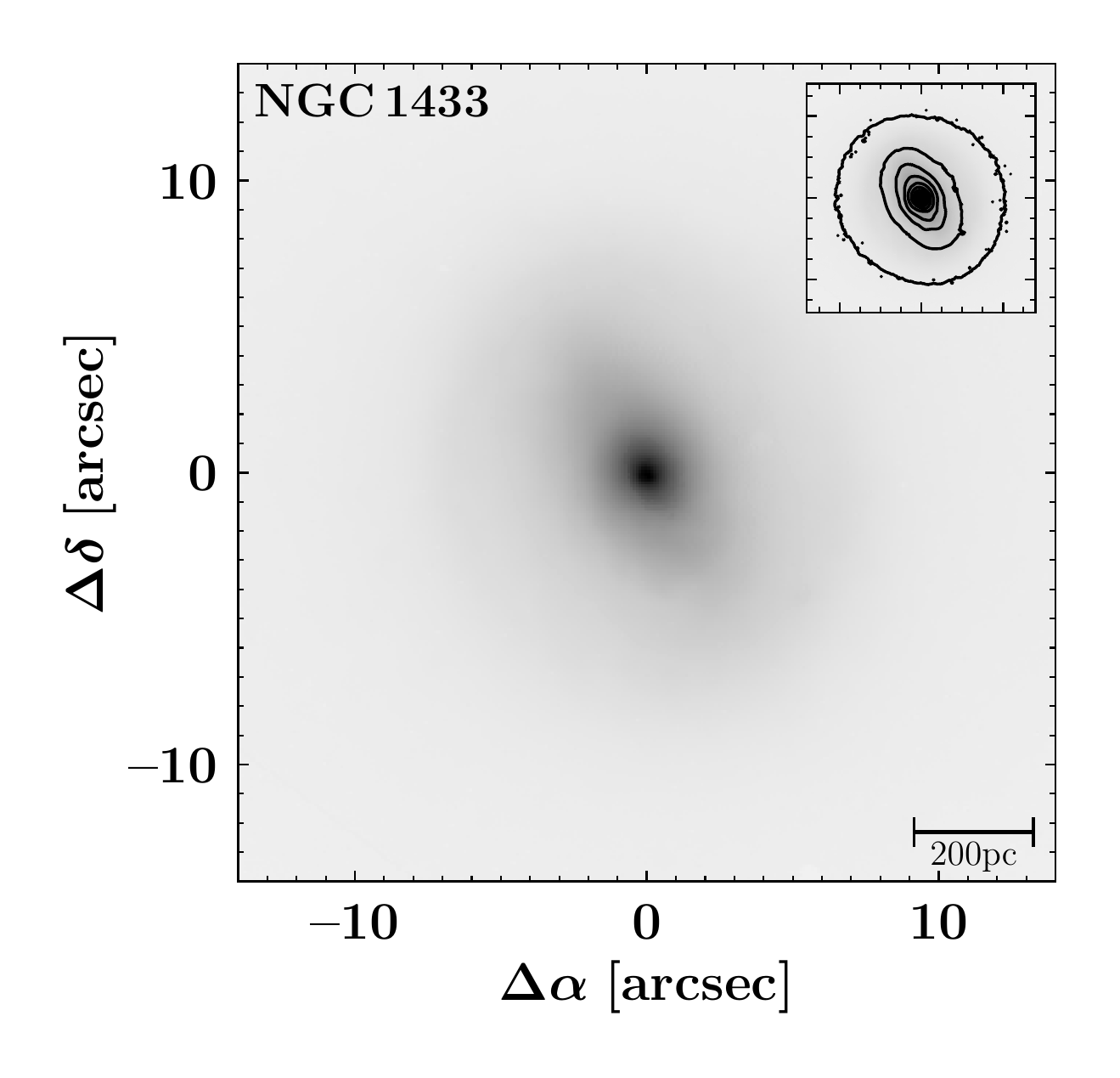}
    \includegraphics[width=0.33\textwidth]{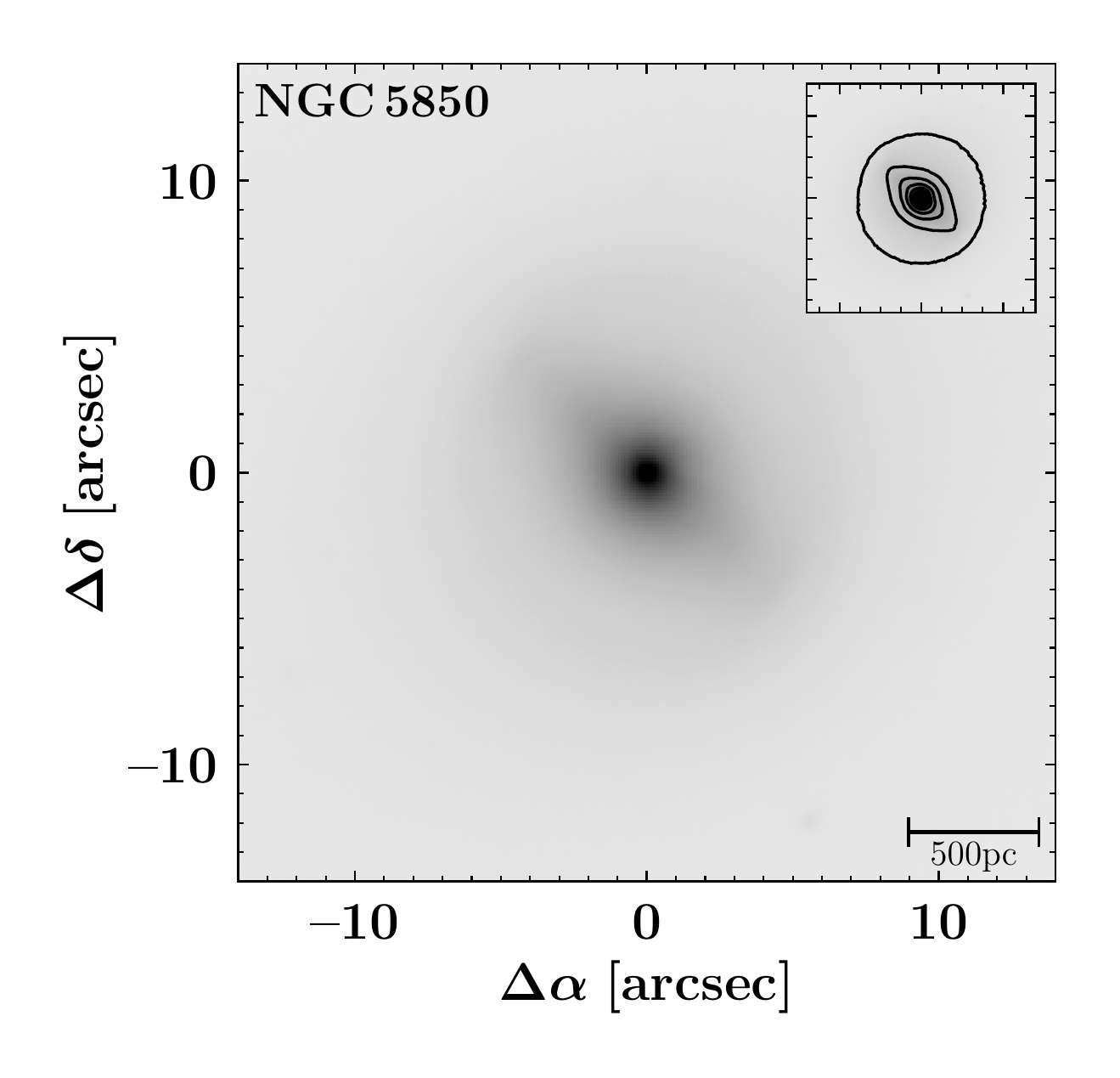}
    \caption{%
        Photometric H-band observations of the inner bars and nuclear discs in NGC\,1291, NGC\,1433, and NGC\,5850,
        obtained with HAWK-I. The insets show exactly the same field of view as the main images and illustrate the
        isophotes in the galaxy centres which highlight not only the axisymmetric structure of the nuclear discs but
        also the significantly elongated inner bars. For the three galaxies the inner bar is clearly distinguishable, in
        particular also the relatively weak inner bar of NGC\,1433 whose existence has been debated. North is up; east
        is to the left. 
    }%
    \label{fig:hawki}
\end{figure*}

In Fig.~\ref{fig:hawki} we present H-band observations of the central regions of the three galaxies, recently obtained
with the HAWK-I imager at ESO's Paranal Observatory. These observations were performed employing the Adaptive Optics
correction produced by the GRAAL AOF module on UT4, leading to a typical PSF of about \SIrange{0.4}{0.5}{\arcsec}. The
high-resolution photometric observations not only highlight the structure of the inner bars in NGC\,1291 and NGC\,5850,
but also show the inner bar in NGC\,1433 more clearly. Although the inner bar in this galaxy does not appear as
prominent as in the two other cases, the isophotes reveal the elongated structure of this inner bar within an almost
axisymmetric nuclear disc. In fact, the detection of inner bars with varying strength is not surprising, as main bars
also exhibit a large range of structural properties. 

In addition, we carefully inspected the inner regions of all other TIMER galaxies using S$^4$G images, our recent HAWK-I
photometry, reconstructed intensities from MUSE, and maps of kinematics and stellar population properties
\citep[][]{bittner2020, gadotti2020}. Although some studies report inner bar fractions of 30\%
\citep[e.g.][]{erwin2002}, no other inner bars are detected in the TIMER sample. In fact, the detected number of inner
bars in the sample is consistent with more recent estimates of a 12\% frequency of double-barred systems (Hildebrandt et
al., subm.). 

In Table~\ref{tab:sampleProperties} we present an overview about some of the fundamental properties of the three
galaxies considered in this study, including inclination, position angle, stellar mass, and distance.  All galaxies
exhibit clear evidence of a nuclear disc and we show their kinematic radii, defined as the radius at which $v/\sigma$
within the nuclear disc reaches its maximum \citep{gadotti2020}.  Finally, we present the basic properties of the inner
bars, i.e.\ their position angles, radii, and axis ratios, based on the visual inspections of \citet{herreraEndoqui2015}
and careful multi-component decompositions of \citet{deLorenzoCaceres2019}. In fact, these decompositions include 6
different galaxy components and thus highlight how complicated the inner regions of these galaxies are. We discuss the
differences in the estimated parameters in Sect.~\ref{subsec:maps}. 

\begin{table*}
    \centering
    \begin{tabular}{lcccccc@{\hspace{1cm}}cccccc}
        \toprule
        \toprule
        Name               &  $i$        &  $M_{\star}$      &  $PA_{gal}$ &  $d$        &  Spatial scale         &  $r_{kin}$  &  Source  &  $PA_{ib}$  &  $a_{full}$    & $a_{vis}$    &  $b/a$     \\
                           &  \si{\deg}  &  \SI{e10}{\msol}  &  \si{\deg}  &  \si{\Mpc}  &  \si{\pc}/\si{\arcsec} &  \si{\kpc}  &          &  \si{\deg}  &  \si{\arcsec}  & \si{\arcsec} &            \\
        (1)                &  (2)        &  (3)              &  (4)        &  (5)        &  (6)                   &  (7)        &  (8)     &  (9)        &  (10)          & (11)         &  (12)      \\
        \midrule                                                                                                             
        NGC\,1291          &  11         &  5.8              &  -8.9       &   8.6       &  42                    &  ---        &  HE      &  18         &  ---           & 19.3         &  0.78      \\
                           &             &                   &             &             &                        &             &  LC      &  17         &  29.0          & ---          &  0.35      \vspace{0.2cm}\\
        NGC\,1433          &  34         &  2.0              &  18.2       &  10.0       &  49                    &  0.381      &  HE      &  30         &  ---           & 7.1          &  0.67      \\
                           &             &                   &             &             &                        &             &  LC      &  ---        &  ---           & ---          &  ---       \vspace{0.2cm}\\
        NGC\,5850          &  39         &  6.0              & -26.5       &  23.1       &  112                   &  0.796      &  HE      &  45         &  ---           & 7.1          &  0.72      \\
                           &             &                   &             &             &                        &             &  LC      &  48         & 11.3           & ---          &  0.21      \\
        \bottomrule                                                                                                                                                
    \end{tabular}
    \caption{%
        Overview of some fundamental properties of the sample. Shown are the inclination $i$ of the galaxy disc relative
        to the plane of the sky in column (2), the total stellar mass $M_{\star}$ in column (3), the position angle of
        the galaxy disc $PA_{gal}$ in column (4) \citep[all from
        S$^4$G;][]{sheth2010,munoz-mateos2013,munoz-mateos2015}, the mean redshift-independent distance $d$ from the
        NASA Extragalactic Database (NED, \url{http://ned.ipac.caltech.edu/}) in column (5), the spatial scale of the
        observations in column (6), and the kinematic radius of the nuclear disc in column (7) \citep{gadotti2020}.
        Column (8) states the source of the structural components described in the following columns: HE refers to the
        results from \citet{herreraEndoqui2015} while LC designates the multi-component decompositions of
        \citet{deLorenzoCaceres2019}. We refer the reader to Sect.~\ref{subsec:maps} for a comparison of the results of
        both methods. Columns (9) to (12) provide the position angle of the inner bar $PA_{ib}$, its full semi-major
        axis $a_{full}$ from photometric decompositions, the inner bar semi-major axis from visual inspections
        $a_{vis}$, and the inner bar axial ratio $b/a$. 
        We note that no kinematic radius could be determined for NGC\,1291, due to the face-on orientation of the
        galaxy. NGC\,1433 is not included in the photometric decompositions of \citet{deLorenzoCaceres2019}. 
    }%
    \label{tab:sampleProperties}
\end{table*}

%
%
%
\subsection{Observations and data reduction}%
\label{subsec:observationReduction}
Exploiting the wide-field mode of the MUSE spectrograph, most TIMER observations were obtained in ESO's period 97 from
April to September 2016. The observations use a wavelength range from \SIrange{4750}{9350}{\angstrom} at a spectral
sampling of \SI{1.25}{\angstrom}. The field of view covers \SI{1}{\arcmin\squared} with a spatial sampling of
\SI{0.2}{\arcsec} at a typical seeing of \SIrange{0.8}{0.9}{\arcsec}. Each observation included approximately 1 hour of
integration on source. Since all galaxies are larger than the field of view, dedicated sky exposures were obtained. 

Following the standard TIMER data reduction procedure, the observations were reduced with version 1.6 of the MUSE data
reduction pipeline \citep{weilbacher2012, weilbacher2020}.  More specifically, the data are calibrated in flux and
wavelength, and bias, flat-fielding, and illumination corrections are applied. Telluric features as well as the sky
background are removed, the latter using principal component analysis. Finally, the data is registered astrometrically.
A detailed overview of observations and data reduction of the TIMER survey is presented in \citet{gadotti2019}. 

%
%
%
\subsection{Data analysis}%
\label{subsec:analysis}
The analysis of the MUSE data is performed within the fully modular software framework of the \texttt{GIST}
pipeline\footnote{\url{http://ascl.net/1907.025}} \citep[Galaxy IFU Spectroscopy Tool,][]{bittner2019}. This tool
provides extensive capabilities for the analysis of spectroscopic data, facilitating all necessary steps from the
read-in and preparation of input data, over its scientific analysis, to the production of publication quality plots.  In
the following, we only summarise the conducted data analysis while a more detailed description is provided in
\citet{bittner2020}. 

We use the \texttt{GIST} framework to exploit the adaptive Voronoi tesselation routine of \citet{cappellari2003}, in
order to spatially bin the data to an approximately constant signal-to-noise ratio of 100. This high signal-to-noise
ratio is employed to ensure the reliability of the measurement \citep[see][for an assessment of
the influence of the signal-to-noise ratio on the derived population properties]{bittner2020}.

We adopt the udf-10 parametrisation of \citet{bacon2017} to model the line-spread function of the MUSE spectra. All
template spectra are broadened to this resolution before any fits are performed.  We further employ the wavelength range
from \SIrange{4800}{5800}{\angstrom} in this analysis. This relatively short portion of the MUSE wavelength range is
chosen because the red part of the spectra is not optimal for the measurement of stellar population properties. In
particular, the lower sensitivity to young stellar populations, residuals from the sky subtraction, as well as 
absorption lines originating in the interstellar medium would complicate the analysis \citep[see e.g.][]{goncalves2020}. 

The measurement of the stellar population properties is performed in three separate steps, using the
full-spectral fitting approach. Firstly, we employ the \texttt{pPXF} routine \citep{cappellari2004, cappellari2017} to
derive the stellar kinematics, while keeping emission lines masked. A low order multiplicative Legendre polynomial is
employed, in order to account for differences in the continuum shape between spectra and templates. Secondly, we apply
\texttt{pyGandALF} \citep{bittner2019}, a new Python implementation of the original \texttt{GandALF} routine
\citep{sarzi2006, jfb2006}, in order to fit any present emission lines while taking into account the results from the
previous stellar kinematics fit. \texttt{pyGandALF} uses a two-component
reddening correction instead of corrective Legendre polynomials that accounts for extinction within the
emission-line regions as well as for ``screen-like'' extinction that affects the entire spectrum. The detection of
emission lines is considered significant if the amplitude-to-residual-noise ratio exceeds four. In these cases, the
emission line is subtracted from the spectrum, thus obtaining emission-subtracted spectra. 

Thirdly, a regularised run of \texttt{pPXF} is performed in order to measure the mean stellar population properties,
using the emission-subtracted spectra. In this process, we apply an 8th order multiplicative Legendre polynomial. We
further fix the stellar kinematics to those derived with the unregularised run of \texttt{pPXF} in order to avoid
degeneracies between metallicity and velocity dispersion \citep[see e.g.][]{sanchezBlazquez2011}.  In fact, \texttt{pPXF}
does not directly derive mean stellar population properties but instead non-parametric star formation histories. This is
achieved by assigning weights to the spectral models so that the observed spectra are best reproduced. However, it is not
straightforward to obtain a physically meaningful solution, as this is an ill-conditioned inverse problem. Therefore,
\texttt{pPXF} applies a regularisation, such that the smoothest solution that is consistent with the data is returned
\citep{press1992, cappellari2017}.  We determine the regularisation strength following the prescription applied, for
instance, by \citet{mcDermid2015} and detailed in \citet{bittner2020}. 

Throughout the analysis, we use the MILES single stellar population (SSP) models \citep{vazdekis2015} as spectral
templates. These models use a Kroupa Revised IMF with a slope of \num{1.30} \citep{kroupa2001} and BaSTI isochrones
\citep{pietrinferni2004, pietrinferni2006, pietrinferni2009, pietrinferni2013}.  The models cover ages from
\SIrange{0.03}{14}{\Gyr}, metallicities from \SIrange{-2.27}{0.40}{\dex}, and {\alphaFe} enhancements of \SI{0.00}{\dex}
and \SI{0.40}{\dex} at a spectral resolution of \SI{2.51}{\angstrom} \citep{jfb2011}. 

In this study, we derive light-weighted stellar population properties, in order to emphasise differences in the stellar
ages. Such light-weighted population properties are obtained by normalising each MILES model by its own mean flux.
Average population properties are then calculated from the weight $w_i$ assigned to the model $i$ via
\begin{eqnarray}
    \langle \mathrm{t}           \rangle =& \dfrac{\sum_i \: w_i \: \mathrm{t}_{\mathrm{SSP, i}}}{\sum_i w_i}  \\
    \langle \mathrm{[M/H]}       \rangle =& \dfrac{\sum_i \: w_i \: \mathrm{[M/H]}_{\mathrm{SSP, i}}}{\sum_i w_i}   \\
    \langle \mathrm{[\alpha/Fe]} \rangle =& \dfrac{\sum_i \: w_i \: \mathrm{[ \alpha /Fe]}_{\mathrm{SSP, i}}}{\sum_i w_i} 
\end{eqnarray}
with the $i$th model having an age of $\mathrm{t}_{\mathrm{SSP, i}}$, metallicity $\mathrm{[M/H]}_{\mathrm{SSP, i}}$,
and an {\alphaFe} enhancement of $\mathrm{[ \alpha /Fe]}_{\mathrm{SSP, i}}$.

These {\alphaFe} enhancements are an indicator of the longevity of star formation in galaxies. As $\alpha$-elements are
predominantly formed in the fusion processes leading to type II supernovae while iron is mostly produced in type Ia
supernovae \citep[see e.g.][]{worthey1992, weiss1995}, the {\alphaFe} abundance gives an estimate of their relative importance. Since
the progenitor stars of type II supernovae are massive and short-lived while those of type Ia supernova have long
lifetimes, the timescale of star formation in a galaxy is constrained by these {\alphaFe} enhancements: High abundances
hint towards short and intense star-formation episodes, while low {\alphaFe} abundances are a signature of slow but
continuous star formation. 

%
%
%

\section{Results}%
\label{sec:results}
In the following we present our results on the stellar population content of inner bars. In particular, we present maps
of the mean population properties and their radial profiles along and perpendicular to the inner bar major axis.  The
measurements of the stellar population properties employed in this study are made publicly
available\footnote{\url{https://www.muse-timer.org}}. 

%
%
%
\subsection{Mean stellar population properties in inner bars}%
\label{subsec:maps}
\begin{figure*}
    \centering{%
        \includegraphics[width=0.99\textwidth]{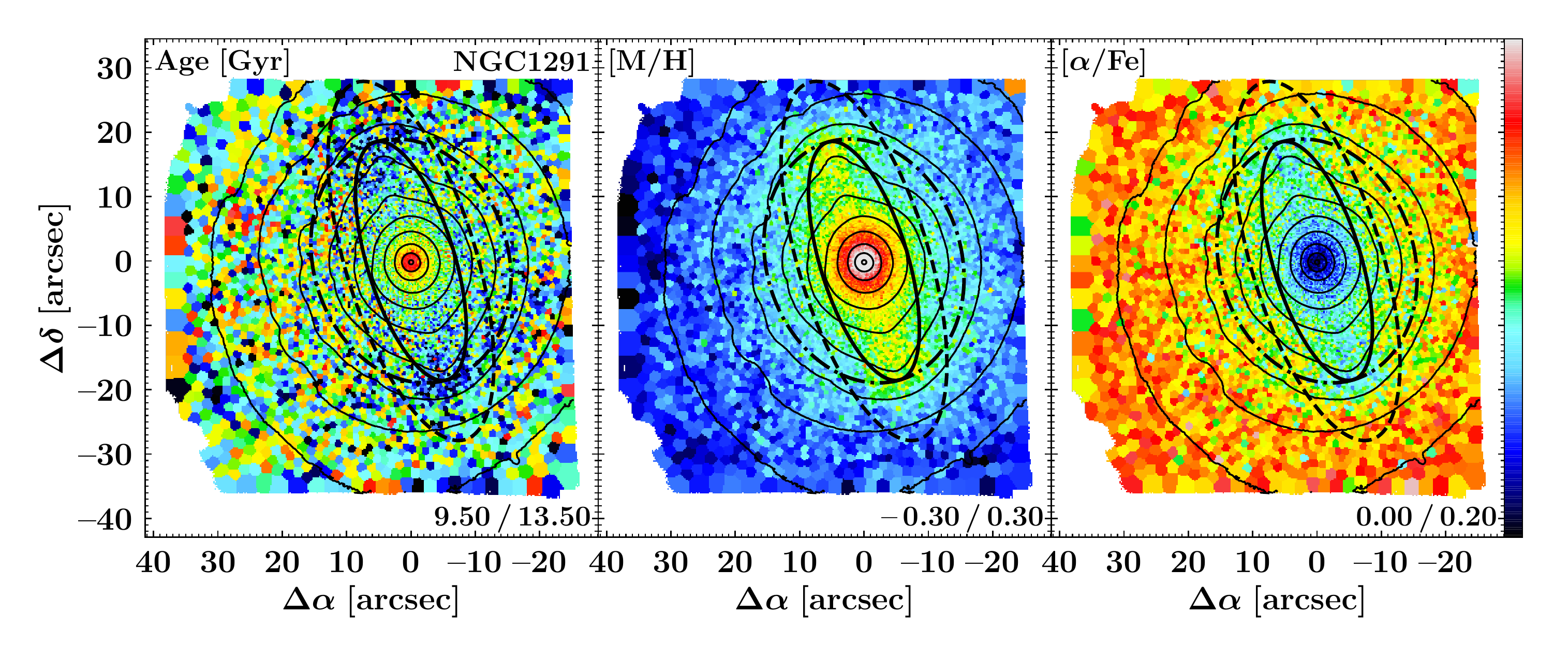} \\
        \includegraphics[width=0.99\textwidth]{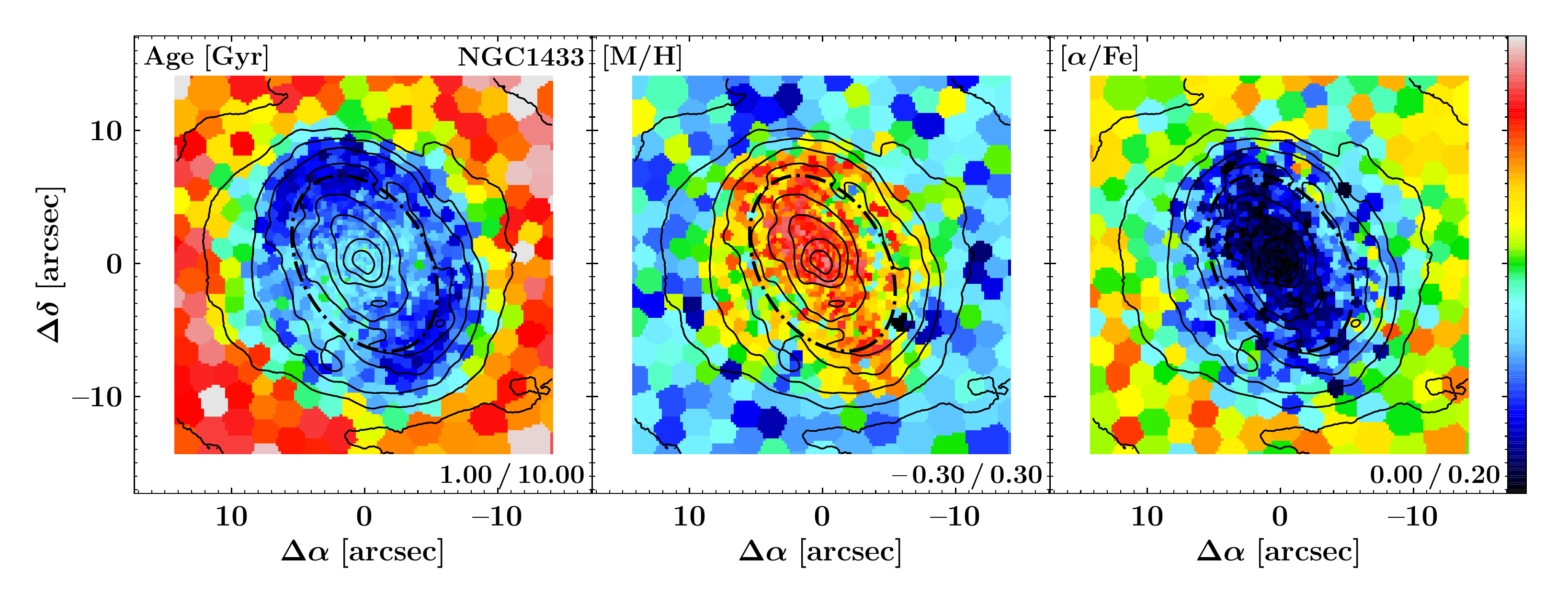} \\
        \includegraphics[width=0.99\textwidth]{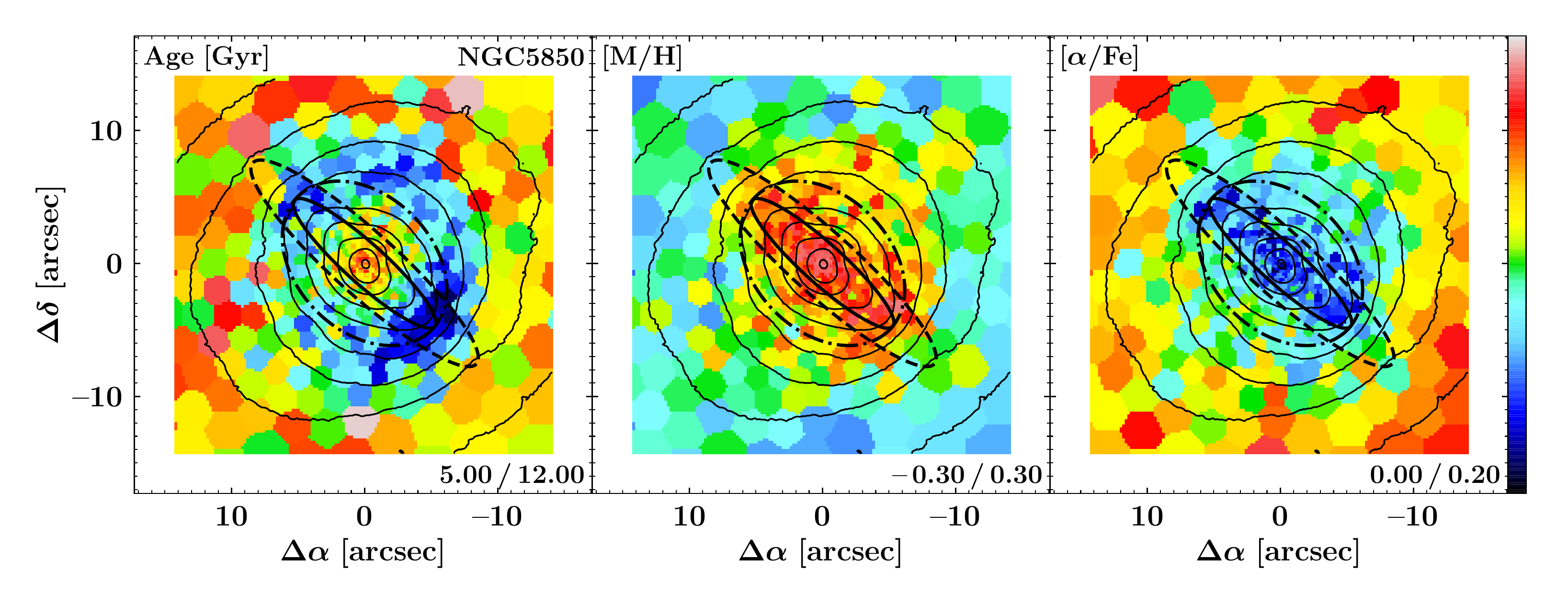}
    }
    \caption{%
        Maps of light-weighted, mean stellar population properties of NGC\,1291 (upper panels), NGC\,1433 (central
        panels), and NGC\,5850 (lower panels), focussed on the spatial region of the inner bar \citep[see][for maps
        showing the entire MUSE field of view]{bittner2020}. Each set of panels displays age, [M/H], and {\alphaFe}
        enhancements and the limits of the respective colour bar are stated in the lower-right corner of each panel.
        The dash-dotted ellipses display the approximate extent of the inner bars according to the visual measurements
        of \citet{herreraEndoqui2015}, while dashed ellipses show the results from the multi-component decompositions of
        \citet{deLorenzoCaceres2019}. Solid ellipses combine the inner bar lengths from \citet{herreraEndoqui2015} with
        the ellipticities of \citet{deLorenzoCaceres2019} and trace the appearance of the inner bars excellently.
        NGC\,1433 is not included in the photometric decompositions of \citep{deLorenzoCaceres2019}, hence only a
        dash-dotted ellipse is displayed.  Based on the reconstructed intensities from the MUSE cube, we display
        isophotes in steps of \SI{0.5}{\mag}. North is up; east is to the left.
    }%
    \label{fig:maps}
\end{figure*}

In Fig.~\ref{fig:maps} we present light-weighted maps of the mean stellar population properties of NGC\,1291, NGC\,1433,
and NGC\,5850, similar to the ones presented in \citet{bittner2020} but focussed on the spatial region of the inner
bar. 
%
In all three galaxies, the inner bars are easily distinguishable, based on their mean stellar populations alone, as
previously claimed by \citet{deLorenzoCaceres2013}. Most
strikingly, the inner bars show significantly elevated metallicities and reduced {\alphaFe} enhancements.  This effect
is most pronounced in NGC\,1291 and NGC\,5850 in which the deviating metallicities and {\alphaFe} abundances are clearly
confined to the long and thin inner bar. To some degree, the same holds for NGC\,1433: while the inner bar in this galaxy appears
slightly rounder, this structure is still distinct in the [M/H] and {\alphaFe} abundance maps. 

In contrast to metallicities and {\alphaFe} enhancements, inner bars do not show significantly different mean stellar
ages in the largest part of their spatial extent. However, in all cases the ends of the inner bars are associated with
relatively younger stellar populations, although these populations can still be very old, especially in the case of
NGC\,1291. These younger ends of the inner bars are broader perpendicular to the inner bar major axis.
Thanks to the superior physical resolution of the observation of NGC\,1291 (simply because this galaxy is closer), this
effect is again most prominent in this galaxy and in line with the morphological properties of its inner bar \citep[see
e.g.][]{mendezAbreu2019, deLorenzoCaceres2019}.

The ends of the inner bars are located close to the outer edge of the nuclear disc, a region that is often highlighted
by gaseous nuclear rings \citep[see][for a discussion of the differences between nuclear discs and nuclear
rings]{bittner2020}. Therefore, the relatively young ends of the inner bar might be affected by recent star formation in
the nuclear ring.  At least for the galaxies in the present sample this is likely not the case. Maps of the {\Ha}
emission-line fluxes which trace HII regions and hence star formation, do not show concentrated emission at the ends of
the inner bars \citep{neumann2020, bittner2020}. Moreover, the galaxies in the present sample do not host gaseous
(star-forming) nuclear rings that might cause the observed effect of inner bars having young stellar populations at
their ends.  In addition, star formation that did not proceed very recently and, thus, might not be traceable with the
{\Ha} emission-line fluxes anymore should not result in such spatially well-defined young ends of the inner bars. Due to
the short dynamical timescale in these radial regions, the young stellar populations would mix and their signature at
the bar ends vanish rapidly. 

Further, we overplot the maps of the mean stellar population properties in Fig.~\ref{fig:maps} with ellipses
highlighting the spatial extent of the inner bars, as previously determined based on photometric data from S$^4$G. 
The black dash-dotted ellipses are based on the visual estimates of the inner bar structural properties from
\citet{herreraEndoqui2015}.  While the length of the inner bar major axis coincides well with the appearance of the
stellar population maps (in particular considering metallicities and {\alphaFe} abundances), the ellipses are in all
cases too round. However, this behaviour is expected, as the measurement is performed on integrated light from multiple
overlapping galaxy components \citep[see also][]{deLorenzoCaceres2020}. In particular the axisymmetric structure of
nuclear discs is expected to bias the inner bar ellipticity towards rounder shapes. 
In contrast, the black dashed ellipses in Fig.~\ref{fig:maps} are derived using detailed, multi-component decompositions
\citep{deLorenzoCaceres2019}. Such photometric decompositions actually separate the light from the different galaxy
components and are, thus, able to better reproduce the elongated shapes of the inner bars observed in the mean population
maps. However, the inner bar full radii are larger than the inner bars appear in the maps. This is a result of the fact
that the bar ends are faint and contaminations from other components, for instance stars in the main bar and main disc, become
significant. 
In fact, these systematic differences of bar lengths and ellipticities derived via different methods have already been
noted in the case of main bars \citep{gadotti2008, gadotti2011}. Taking into account these differences, 
in the remainder of this paper we use inner bar lengths from
the visual estimates of \citet{herreraEndoqui2015} and ellipticities from the multi-component decompositions of
\citet{deLorenzoCaceres2019} in order to highlight the spatial extent of inner bars (solid ellipses in Fig.~\ref{fig:maps}).
This combination of photometric estimates traces the appearance of inner bars in the stellar population maps
excellently.  Nonetheless, the differences between the measurement methods again show how complicated the central
regions of these galaxies are due to multiple, overlapping
galaxy components. 

The stellar population properties of the inner bar of NGC\,5850 were first studied by
\citet{deLorenzoCaceres2013}, based on observations with the SAURON spectrograph \citep{bacon2001}. To this end, they
derived single stellar population equivalent population properties from the measurement of line strength indices at a
signal-to-noise level of 60. In agreement with the results presented here, they find that the inner bar of NGC\,5850 has
systematically higher metallicities compared to its immediate surroundings. However, in their study the inner bar shows
only a weak indication of systematically lower {\alphaFe} abundances. We speculate that this is a result of the different
measurement methods, i.e.\ the use of line strength indices to infer an [Mg/Fe] overabundance, as compared to the use of
full spectral fitting to derive {\alphaFe} enhancements here. 

In a previous TIMER publication, \citet{deLorenzoCaceres2019} investigated various possible formation scenarios for
inner bars based on an independent analysis. In their study, they choose to spatially bin the MUSE data to a
signal-to-noise ratio of 40 (as compared to a signal-to-noise of 100 used here), model and remove emission lines with
\texttt{GandALF}, and fit ages and metallicities with {\steckmap} \citep{ocvirk2006a, ocvirk2006b}. {\alphaFe}
enhancements are estimated based on a set of line-strength indices. While their analysis uses different techniques and
routines, the stellar population properties agree qualitatively well with the results presented here, in particular when
considering that {\ppxf} returns systematically higher stellar ages than {\steckmap} \citep[see][]{bittner2020}.  In
fact, the stellar population properties of the inner bar of NGC\,5850 agree well with the ones presented here.
\citet{deLorenzoCaceres2019} conclude that the middle part of the inner bar in NGC1291 is slightly older compared to its
surroundings, while in the present study no age difference is detected in this part of the inner bar. In fact, as
{\steckmap} returns systematically younger stellar ages, subtle differences in age are emphasised that remain unnoticed
with {\ppxf}, especially at the observed ages of \SI{>10}{\Gyr} at which it is challenging to reliably
distinguish stellar ages. Nonetheless, a careful visual comparison reveals that the young ends of the inner bar appear
to be already detected by \citet{deLorenzoCaceres2019}. Hence, the stellar population properties estimated in the two
studies are consistent. 

%
%
%
%
\subsection{Profiles along the inner bar major and minor axis}%
\label{subsec:radialProfiles}
\begin{figure*}
    \centering{%
        \includegraphics[width=0.33\textwidth]{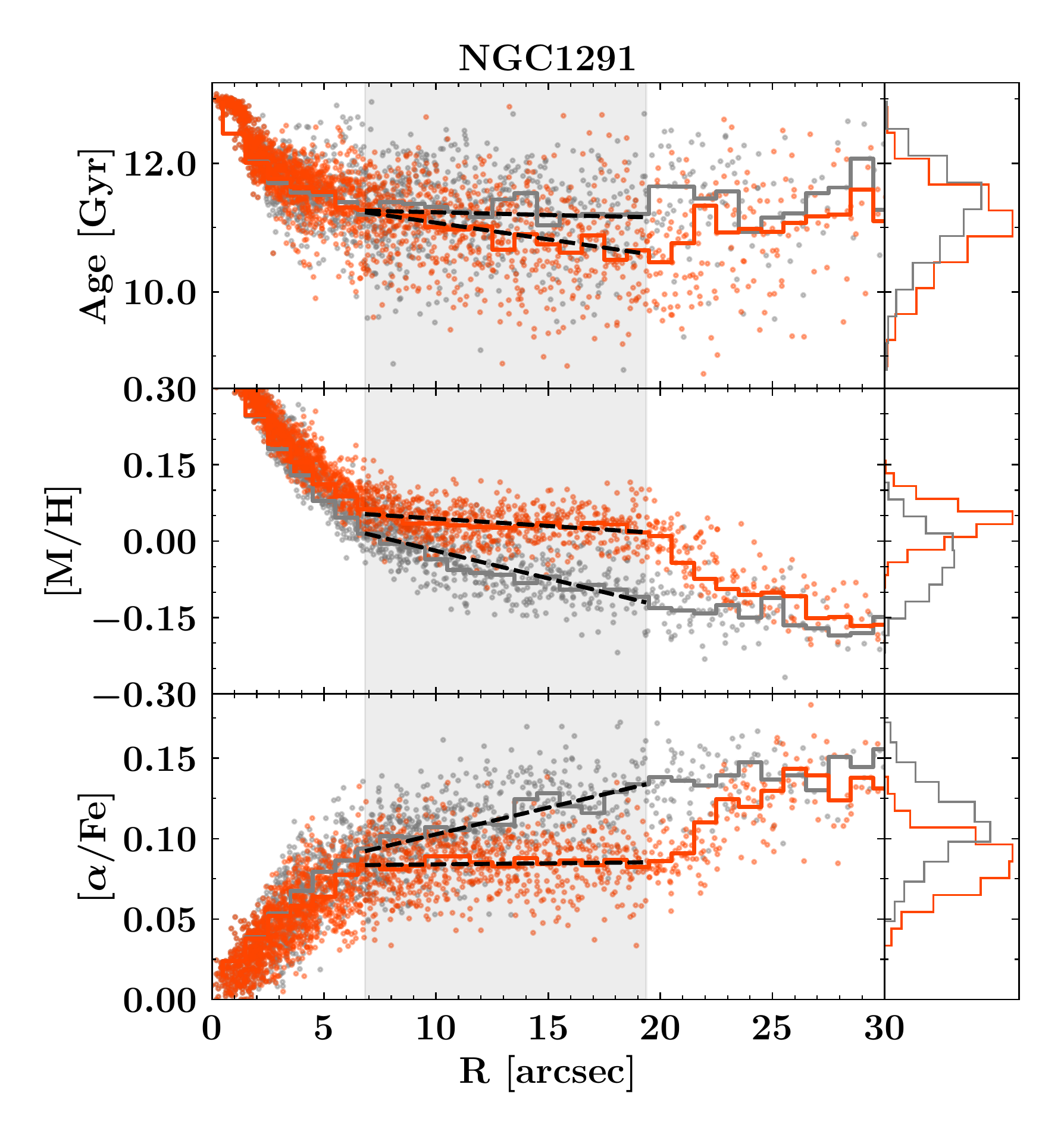}%
        \includegraphics[width=0.33\textwidth]{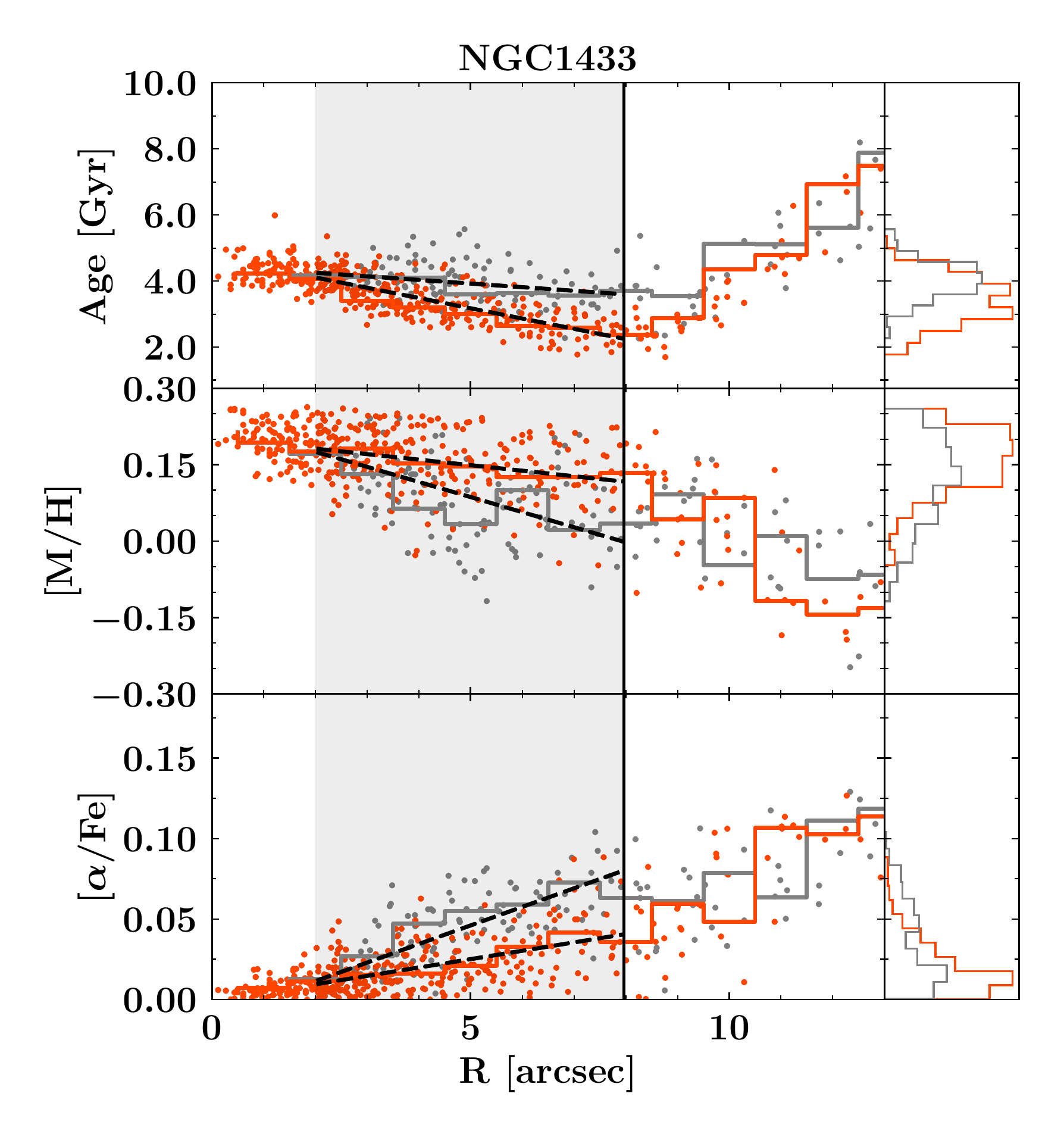}%
        \includegraphics[width=0.33\textwidth]{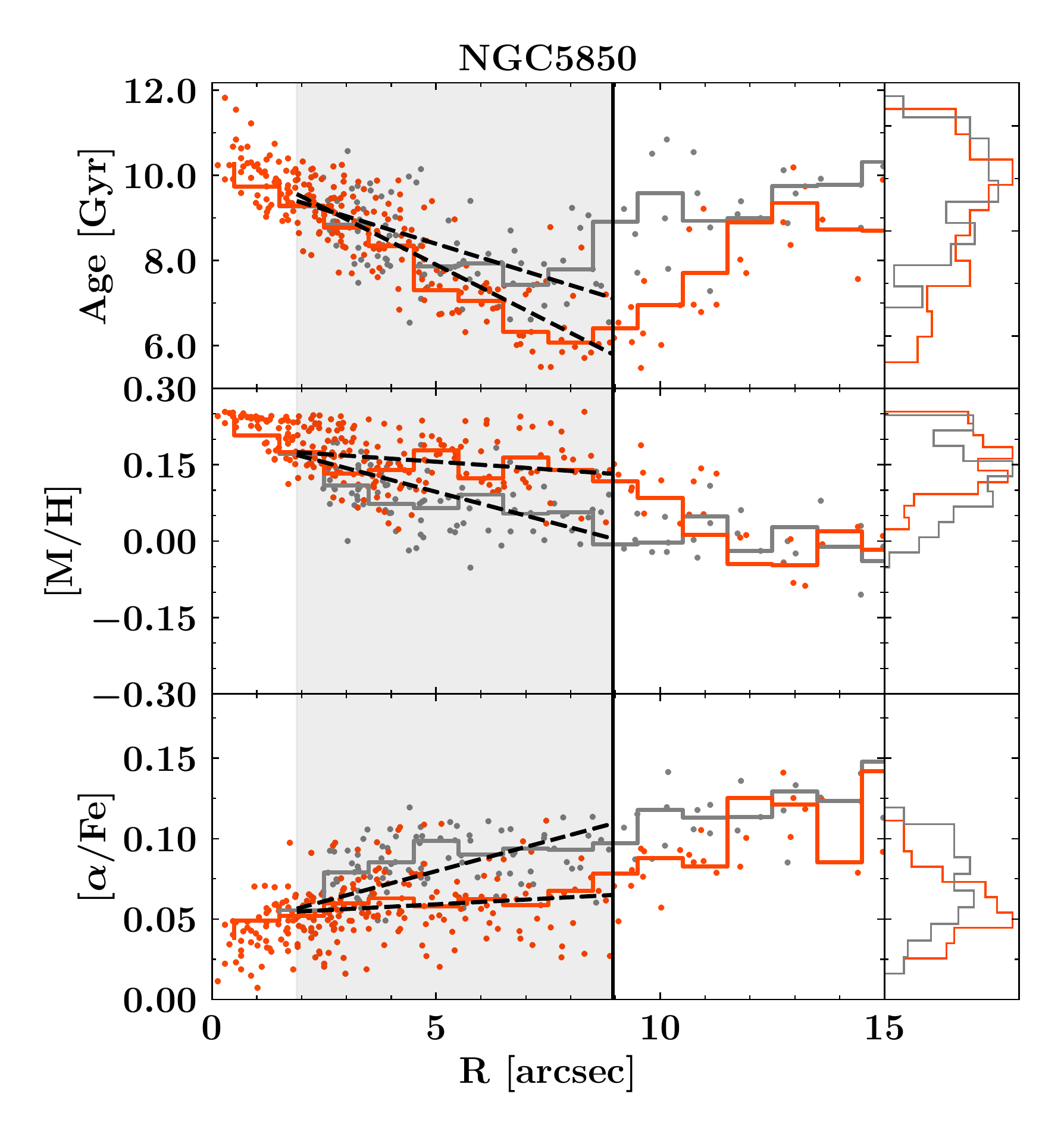}
    }
    \caption{%
        Radial profiles of ages (upper panels), metallicities (central panels), and {\alphaFe} abundances (lower panels)
        as a function of the galactocentric radius. The profiles have been deprojected using inclinations and position
        angles derived in S$^4$G, as presented in Table~\ref{tab:sampleProperties}. Profiles along and perpendicular to
        the major axis of the inner bar are displayed in orange and grey, respectively.  To better highlight the radial
        trends, we also plot the median population properties in bins of \SI{1}{\arcsec} (solid lines). 
        Shaded regions highlight the radial range of the inner bars, i.e.\ for NGC\,1291 and NGC\,5850 the range between
        the inner bar minor and major axis radius, calculated using the inner bar radius from \citet{herreraEndoqui2015}
        and ellipticity from \citet{deLorenzoCaceres2019}. Since the inner bar properties of NGC\,1433 are not
        constrained by photometric studies, we highlight the radial range between \SI{2}{\arcsec} (in order to
        approximately exclude the range in which both the minor and major axis profiles probe the inner bar) and the
        kinematic radius of the nuclear disc from \citet{gadotti2020}. 
        Dashed lines show linear fits to the population profiles within the radial range of the inner bar. The
        histograms on the right display the distribution of stellar population properties in the inner bar major and
        minor axis, again within the radial range of the inner bar. 
        The vertical solid lines represent the kinematic radii of the nuclear discs \citep{gadotti2020}.  We note that
        for NGC\,1291 no kinematic radius could be determined, due to the face-on orientation of the galaxy. 
    }%
    \label{fig:profiles}
\end{figure*}

\begin{figure}
    \centering
    \includegraphics[width=0.5\textwidth]{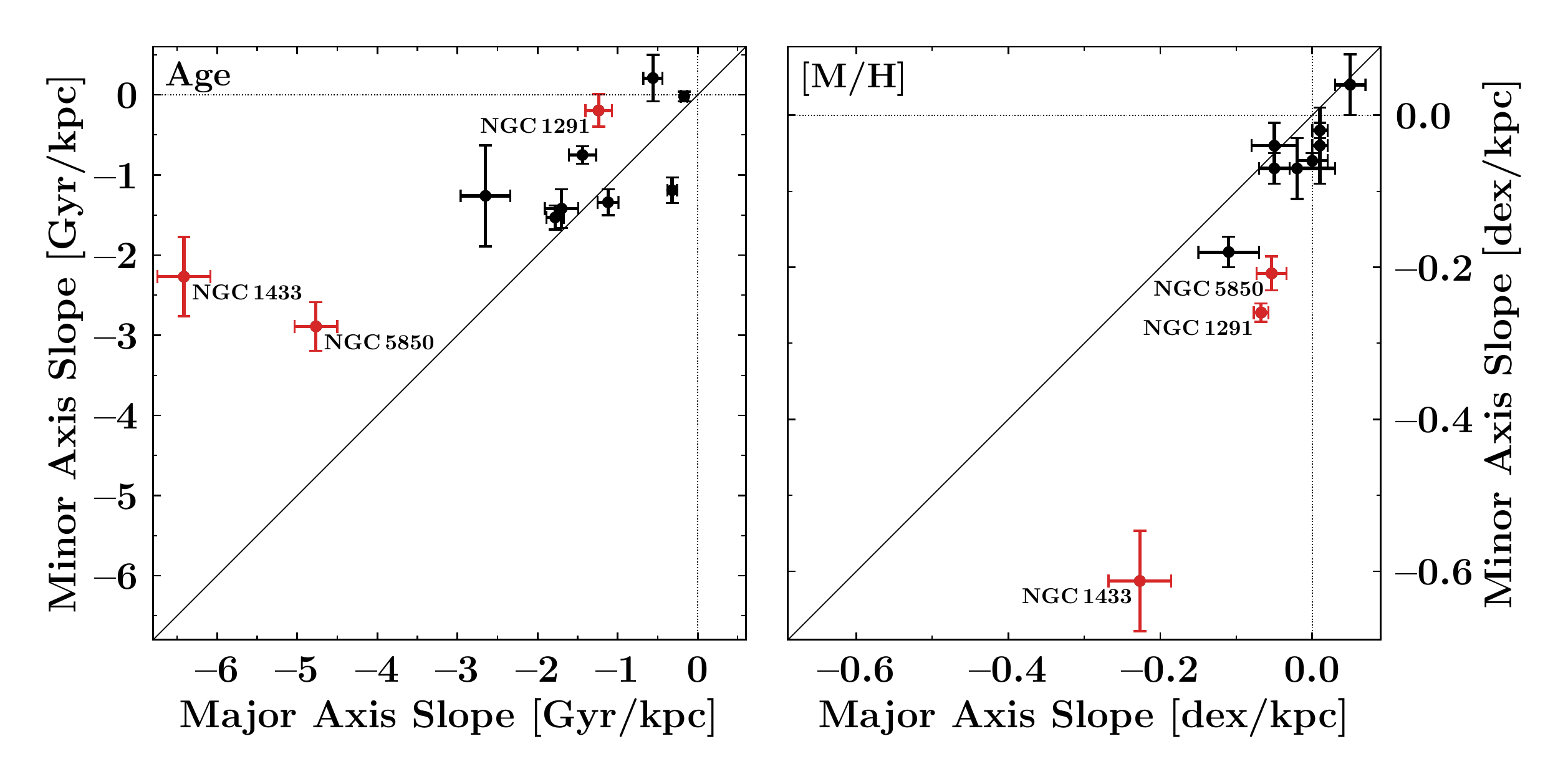}
    \caption{%
        Comparison of the age (left-hand panel) and metallicity (right-hand panel) slopes measured along the major and
        minor axis of the inner bars in this study (red data) and the main bars from \citet[][black data]{neumann2020}. 
    }%
    \label{fig:barSlopes}
\end{figure}

\begin{table*}
    \centering
    \begin{tabular}{llrrrrrr}
        \toprule
        \toprule
                   &      &  age-slope            & [M/H]-slope         & {\alphaFe}-slope  & <Age>       & <[M/H]>     & <{\alphaFe}> \\
        \midrule
        NGC\,1291  &  MA  &  \num{-5.15(69) e-2}  & \num{-2.8(4) e-3}   & \num{1(1) e-4}    & \num{10.97} & \num{0.04}  & \num{0.08}   \\\vspace{0.2cm}
                   &  MI  &  \num{-8.1(84) e-3}   & \num{-1.08(5) e-2}  & \num{3.3(3) e-3}  & \num{11.22} & \num{-0.04} & \num{0.11}   \\
        NGC\,1433  &  MA  &  \num{-3.11(16) e-1}  & \num{-1.1(02) e-2}  & \num{5.2(5) e-3}  & \num{3.43}  & \num{0.16}  & \num{0.02}   \\\vspace{0.2cm}
                   &  MI  &  \num{-1.10(24) e-1}  & \num{-2.97(32) e-2} & \num{11.5(07) e-3}& \num{4.03}  & \num{0.12}  & \num{0.03}   \\
        NGC\,5850  &  MA  & \num{-5.338(297) e-1} & \num{-6.0(22) e-3}  & \num{1.5(8) e-3}  & \num{8.28}  & \num{0.16}  & \num{0.06}   \\
                   &  MI  & \num{-3.236(339) e-1} & \num{-2.33(25) e-2} & \num{7.5(9) e-3}  & \num{8.73}  & \num{0.12}  & \num{0.07}   \\
        \bottomrule
    \end{tabular}
    \caption{%
        Slopes of the linear fits to the stellar population properties along the inner bar major (MA) and minor axis
        (MI) in the radial range of the inner bar (in units of \si{\Gyr\per\arcsec} and \si{\dex\per\arcsec}). We
        further provide the mean stellar population properties in this radial range (in units of \si{\Gyr} and
        \si{\dex}). 
    }%
    \label{tab:values}
\end{table*}

In Fig.~\ref{fig:profiles} we plot light-weighted stellar ages, metallicities, and
{\alphaFe} abundances as a function of the deprojected galactocentric radius.  These profiles are calculated based on
pseudo-slits crossing the galaxy centre and being oriented along and perpendicular to the major axis of the inner bar,
according to the position angles presented in Table~\ref{tab:sampleProperties}.  We include all spatial bins within a
pseudo-slit with a width of \SI{2}{\arcsec} but check that using different widths, e.g. \SI{1}{\arcsec} or
\SI{4}{\arcsec}, do not alter the obtained profiles qualitatively. 

Naturally, the radial population profiles clearly corroborate the findings from the stellar population maps.  In
Table~\ref{tab:values} we provide an overview of the mean stellar population properties measured along and perpendicular
to the inner bar major axis. Average differences vary between galaxies, but indicate that inner bars are
systematically more metal-rich and less {\alphaFe} enriched (see Table~\ref{tab:values}).  We note that the differences
between the major and minor axis depend on the radius and, hence, the typical differences at the inner bar radius are
significantly larger than the average differences in the radial region of the inner bar.  Considering stellar ages,
inner bars seem to be slightly younger, although typical age differences are relatively small. In line with our findings
from the population maps, the largest age differences are found close to or just outside of the inner bar radius,
reaching maximum age differences of \SI{2}{\Gyr} for NGC\,5850, and between \SIrange{1}{2}{\Gyr} in the case of
NGC\,1433.  These absolute age differences at the ends of the inner bar remain below \SI{1}{\Gyr} only for NGC\,1291. In
fact, the observed age differences somewhat depend on the absolute values of stellar age. Distinguishing
the ages of stellar populations at old ages, for instance in the case of the \SI{>10}{\Gyr} populations in NGC\,1291, is
very difficult, possibly indicating that in this case the age differences are not significant.

Another interesting aspect are the slopes of the stellar population profiles inside the inner bars, especially when
considering metallicity and {\alphaFe} abundances. The [M/H] and {\alphaFe} profiles along the inner bar major axis of
NGC\,1291 are steeply decreasing/increasing from the centre, an effect that is most likely due to the small
kinematically hot spheroid in this galaxy with an effective radius of \SI{9.9}{\arcsec} \citep{deLorenzoCaceres2019}.
However, the profiles are remarkably flat in the radial range occupied by the inner bar (shaded area in
Fig.~\ref{fig:profiles}). A prominent break in the profiles is evident at the inner bar radius where [M/H] and
{\alphaFe} steeply decrease/increase and converge to the lower/higher values typically observed outside of the nuclear
disc. In contrast, the slopes along the inner bar minor axis differ significantly from those along the inner bar major
axis. In fact, the minor axis profiles smoothly connect the steep slopes in the innermost region to the values at large
radii, showing a larger/lower slope. The inner bar of NGC\,5850 follows exactly the same trends, however, the
effect is less prominent for NGC\,1433. The latter is not surprising, as the inner bar in this galaxy is somewhat weaker
and less elongated. 
Interestingly, such flat radial profiles are not evident when considering the stellar ages of any galaxy in the sample.
Instead, the observed stellar ages are continuously decreasing both along and perpendicular to the inner bar. In all
three galaxies of the present sample, the stellar ages are decreasing more strongly along the inner bar major axis than
perpendicular to it.

In order to better quantify these findings, we fit ages, metallicities, and {\alphaFe} abundances of the three galaxies
in the radial range of the inner bar (shaded area in Fig.~\ref{fig:profiles}). For NGC\,1291 and NGC\,5850 this is the
radial range between the inner bar minor and major axis radius, as determined from the measurements of
\citet{deLorenzoCaceres2019} and \citet{herreraEndoqui2015}. Equivalently, for NGC\,1433 
we choose the radial range between \SI{2}{\arcsec} (in order to exclude the radial
region where both major and minor axis still probe the inner bar) and the kinematic radius of the nuclear disc from
\citep{gadotti2020}. The resulting fits are shown in Fig.~\ref{fig:profiles} and the results summarised in
Table~\ref{tab:values}.  These measurements support the considerations above and confirm that the radial profiles of
metallicity and {\alphaFe} are significantly flatter along the inner bar major axis, as compared to the minor axis
profiles. 

In Fig.~\ref{fig:barSlopes}, we compare the absolute values of the age and metallicity slopes along the inner bar major
and minor axis with those obtained by \citet{neumann2020} for main bars in different TIMER galaxies. The relative
steepening/flattening of the age/metallicity profiles along the inner bar major axis -- as compared to the minor axis
-- is well compatible with the results
for main bars, as indicated by the systematic offset from the one-to-one correspondence line. The absolute values of the
slopes of the inner bars may vary substantially from those of main bars but this is expected from the different assembly
history of nuclear discs and main galaxy discs. In contrast to main discs, nuclear discs are formed through bar-driven
processes \citep{bittner2020}, hence the absolute age slope is set by the timescale of these processes. Similarly, a
comparison of minor axis fits between inner and main bars is elusive, as those probe distinct structural components of
the galaxies, i.e.\ nuclear discs and main discs.   

Finally, we point out that the relative differences in the stellar population properties discussed above are of the same
order of magnitude as the typical measurement errors presented in the literature \citep[see e.g.][]{gadotti2019,
pinna2019, bittner2020}.  However, while comparing absolute values of stellar population properties between
different galaxies involves large uncertainties, this is not the case when considering trends within the same galaxy. In
other words, it is unlikely that the observed systematic differences in the population content within one galaxy arise
from random uncertainties in the measurement procedure, in particular since the trends presented in
Fig.~\ref{fig:profiles} are averaged over many spatial bins. In the right-hand panels of
Fig.~\ref{fig:profiles} we compare the distributions of stellar population properties along and perpendicular to the
inner bar major axis within its radial range. For ages, metallicities, and {\alphaFe} abundances of both galaxies, an
Anderson-Darling test confirms at a \SI{99.9}{\percent} significance level that both observed distributions are not
drawn from an identical parent distribution. 

%
%
%

\section{Discussion}%
\label{sec:discussion}
In this section we discuss our findings on the stellar population properties of inner bars in comparison with
analogous measurements in main bars. We further highlight these results in the context of the dynamical mechanisms of
orbital age separation and orbital mixing. We remind the reader that the galaxies in the present sample exhibit high total
stellar masses (see Table~\ref{tab:sampleProperties}) and low-mass barred galaxies might follow trends different from
those discussed in the following \citep[see e.g.][]{elmegreen1985, kruk2018, zurita2020}. 

%
%
%
\subsection{Stellar populations in bars, inner bars, and their star formation deserts}%
\label{subsec:meanStellarPopulations}
Main bars are not only detectable based on their morphology or kinematic features, but can also be identified by their
stellar population content. In particular, bars typically exhibit higher metallicities and show lower {\alphaFe}
abundances, as compared to their immediate surroundings or the main disc \citep[see e.g.][]{perez2007, perez2009,
sanchezBlazquez2011, sanchezBlazquez2014, fragkoudi2020, neumann2020}. While this finding does not necessarily hold true
when comparing bars to the outer parts of the main galaxy disc \citep[see e.g.][]{seidel2016}, it does when bars are
compared to the part of the disc within the bar radius but outside of the bar itself, a region often denoted as ``star
formation desert'' \citep[SFD; see e.g.][]{james2009, james2016, donohoeKeyes2019}. Basically, these studies suggest
that star formation in the SFD is suppressed very rapidly after bar formation. In contrast, star formation in the bar
itself continues at least for a limited amount of time, although at relatively low rates, resulting in slightly younger
stellar populations in the bar. Due to this continuing self-enrichment, the bar eventually reaches higher metallicities
and depleted {\alphaFe} enhancements, as compared to the SFD\@. Hence, the star formation desert can be identified as a
region within the bar radius but outside of the bar itself that is characterised by relatively low metallicities and
elevated {\alphaFe} abundances. 

Interestingly, in Sect.~\ref{subsec:maps} we find similar results for the stellar population content of inner bars. In
particular, inner bars exhibit elevated values of [M/H] and depleted {\alphaFe} abundances, as compared to the region
outside of the inner bar but within the inner bar radius (i.e.\ the star formation desert of the inner bar). Moreover,
Fig.~\ref{fig:profiles} indicates that inner bars are slightly younger along than perpendicular to their major axis.
Nonetheless, previous findings \citep[e.g.][]{deLorenzoCaceres2019} indicating that the inner bar of NGC\,1291 is
slightly older compared to its SFD suggest that, at least in some cases, star formation is suppressed more
rapidly in the bar itself \citep[see also][]{verley2007, neumann2019, fraserMcKelvie2020, diaz-garcia2020, neumann2020}. 

In general, based on maps of the mean stellar population properties alone, bars and inner bars appear similar in terms
of their relative population content. To date, inner bars were only marginally resolved in integral-field spectroscopic
observations, due to the different spatial scales of main and inner bars.  However, this limitation is now overcome by
the high spatial sampling of the MUSE spectrograph. Visually comparing the inner bars in this study to the main bars
observed in the TIMER survey \citep[see][]{neumann2020}, highlights the similarities of the stellar population content
between bars and inner bars. In addition, a comparison of the properties of inner bars to the area within the inner bar
radius but outside of the inner bar itself suggests the presence of a star formation desert related to inner bars, again
as in main bars. 

%
%
%
\subsection{Flat population gradients along inner bars: orbital mixing}%
\label{subsec:orbitalMixing}
Another relevant aspect when considering mean stellar population properties of bars are their gradients along the bar
major axis, an aspect that holds information on their dynamical structure. More specifically, bars are elongated
components which are built-up by a large number of elongated stellar orbits. These orbits might be composed of different
stellar populations, show distinct levels of elongation, and reach different maximum radii in the bar.  However, this
does not only result in more elongated orbits dominating the spatial regions at the ends of the bars (see
Sect.~\ref{subsec:orbitalAgeSeparation}), but also causes different orbits to come very close to each other within the
majority of the radial extent of the bar. In other words, stars on different orbits cross the same spatial region in the
galaxy \citep[see e.g.][]{binney1987, contopoulos1989, athanassoula1992a}. Therefore, at each location in the bar the
light contributions of different stellar populations on different orbits appear mixed and the gradients of the stellar
population content in the main parts of the bars, in particular [M/H] and {\alphaFe}, are expected to be comparably
flat. 

In fact, recent observations suggest a flattening of the stellar populations gradients along the major axis of
the bar, well consistent with the theoretical considerations above \citep[see e.g.][]{sanchezBlazquez2011, williams2012,
seidel2016, fraser-mcKelvie2019, neumann2020}. To date, radial population gradients could only be derived for main bars,
as the spatial resolution remained a limiting factor. 

In this study, we present for the first time radial gradients of the mean stellar population properties along both the inner
bar major and minor axis.  Similar to main bars, we find that the radial gradients of metallicities and {\alphaFe}
abundances along the inner bars major axis are significantly flatter compared to the profiles along the minor axis.
This flattening resembles the stellar population profiles previously observed in main bars. This suggests that inner
bars have a significant impact on the radial distribution of stellar populations in nuclear discs and hence that the
orbital mixing which is evident in bars occurs in inner bars as well.  Again, this result reinforces the idea that inner
bars are dynamically distinct components that form and evolve in the same way main bars do. 

Interestingly, while a flattening of the radial gradients along the inner bar major axis is prominent for metallicities
and {\alphaFe} abundances, no such gradient is evident in the stellar age profiles. Instead, the measured stellar ages
typically decrease from the centre of the galaxy to the outer edge of the nuclear disc.  In fact, such negative
gradients in age are frequently observed in nuclear discs, regardless of the presence of an inner bar, and appear to be
a result of their inside-out formation scenario \citep{bittner2020}.  In addition, orbital age separation, i.e.\ the
dynamical effect that the ends of bars are younger than their central parts, should also hinder the development of
flat age profiles. We detail this aspect in the following subsection.

%
%
%
\subsection{Young ends of inner bars: orbital age separation}%
\label{subsec:orbitalAgeSeparation}
Main bars do not only appear distinguished from their surroundings by their elevated metallicities and depleted
{\alphaFe} abundances, but also show spatially well-defined variations in age. The ends of main bars generally appear
significantly younger compared to the rest of the bar. While this effect could be a result of enhanced star formation in
these regions, the inspection of {\Ha} emission-line maps has shown that no signatures of ongoing star formation are
detected. Instead, this observation is a result of orbital age separation (also referred to as kinematic fractionation)
and known from both observational and numerical studies on the stellar population content of bars \citep[see
e.g.][]{perez2007, wozniak2007, fragkoudi2017, fragkoudi2018, athanassoula2017, debattista2017, neumann2020}. 

These studies find that distinct kinematic components of the galaxy disc participate in bars in different ways.
Kinematically cold stars (i.e.\ those with a low radial velocity dispersion) preferably end up in highly elongated $x_1$
orbits (oriented along the bar major axis) while kinematically hot stars (i.e.\ those with high radial velocity
dispersion) form rounder and less elongated bar $x_1$ orbits. Since more elongated orbits extend to larger radii, the
ends of the bars are dominated by stars on the most elongated orbits.  In the main disc of a typical galaxy, these
kinematically cold and hot components could be the thin and thick disc \citep{fragkoudi2017, debattista2017,
athanassoula2017}. While the thick disc is typically thought to be old, metal-poor, and shows high velocity dispersions,
the thin disc is younger, more metal-rich, and exhibits low velocity dispersions \citep[see e.g.][]{prochaska2000,
cheng2012, pinna2019}. As a consequence of this, the young stars of the thin disc end up on more elongated bar orbits
and thus dominate the spatial region at the ends of the bar, hence creating the effect that the ends of bars show
younger stellar populations. 
Interestingly, this effect is most prominent for stellar ages, but absent when considering metallicities and {\alphaFe}
abundances, in contrast to the simulations of \citet{fragkoudi2018}. A possible explanation could be that age is a
better tracer of the stellar kinematics, or, more precisely, of the velocity dispersion \citep[see e.g.][]{peletier2007,
jfb2016}. The longer a stellar population was present in the disc of a galaxy, the longer it was exposed to dynamical
heating processes, resulting in an increased velocity dispersion.  Hence, the velocity dispersion should be a function
of stellar age and, therefore, the ends of bars can be observed to exhibit younger stellar populations.  In contrast,
stars of a given age can have a range of metallicities and {\alphaFe} enhancements, so that these quantities are not
necessarily a function of stellar velocity dispersion, and therefore the ends of bars and inner bars do not necessarily
show distinct [M/H] and {\alphaFe} values.

In all galaxies in this study, we do detect such younger stellar populations in a spatially confined region at the
ends of the inner bars. As a result, these relatively young ends spatially coincide with $\sigma$-hollows in these
galaxies \citep{deLorenzoCaceres2019, mendezAbreu2019, bittner2019}. These regions of significantly lower stellar
velocity dispersion are typically detected towards the ends of inner bars, i.e.\ in areas in which the inner bar
dominates the stellar light \citep{deLorenzoCaceres2008, deLorenzoCaceres2013}. Observing younger stellar populations in
the same regions hints towards the existence of the dynamical effect of orbital age separation in inner bars. This again
suggests that inner bars work dynamically identical to bars, differing only in the spatial scale on which they form from
disc instabilities. 

This result also indicates that stars from kinematically different components participate in the inner bar.
Analogous to the main bar existing in both the thick and thin main disc, one might speculate if nuclear discs consist of
hotter and colder nuclear disc components as well or that the hot main disc component might participate in the inner
bar.  However, dedicated studies, in particular numerical simulations of nuclear disc formation in a self-consistent
cosmological context, are necessary to shed further light on these aspects. 

Finally, a frequent photometric feature of strong bars in early-type spirals are surface brightness enhancements at
their ends, a structure typically referred to as \textsl{ansae} \citep{martinezValpuesta2007, buta2015}. To date, these
structures have not been studied intensively and their physical origin as well as presence in inner bars remains
elusive.  Here we speculate that the systematically younger stellar populations at the ends of inner and main bars could
be connected to photometric ansae. In fact, these younger stellar populations, in particular in the case of NGC\,1291,
resemble the morphological appearance of ansae in photometric studies. Such younger stellar populations are expected to
be relatively brighter and (at a given surface mass density) should result in elevated surface brightnesses.  Other
mechanisms, such as enhanced star formation at the ends of bars, or the dynamical trapping of stars around the ansae
\citep{martinezValpuesta2007}, might as well contribute to this phenomenon. 

%
%
%

\section{Summary and conclusions}%
\label{sec:summary}
We have exploited MUSE observations of the central regions of the three galaxies NGC\,1291, NGC\,1433, and NGC\,5850,
observed as part of the TIMER survey. All galaxies clearly exhibit nuclear discs, and two of them host prominent inner
bars. The inner bar in NGC\,1433 appears weaker and less elongated particularly in earlier imaging data, but is clearly
detected in our new, AO-assisted, H-band photometric
observations. In the present study, we have used the full spectral fitting code {\ppxf} in order to derive spatially
resolved maps and radial profiles of the mean stellar population properties in the inner bars.  We compare these results
to those obtained for main bars, and further use them to constrain their dynamical structure. Our main results are as
follows: 

\textbf{(i)}
Inner bars can be clearly distinguished, based on their mean stellar population properties alone. In particular, they
are characterised by elevated metallicities and depleted {\alphaFe} abundances, as compared to the region outside of the
inner bar but within the inner bar radius. Although inner bars show systematically younger ages, these differences are
generally small. Based on spatially resolved, high-resolution maps of the mean population properties and their relative
values, inner and main bars appear identical. 

\textbf{(ii)}
Radial profiles of metallicities and {\alphaFe} abundances along the inner bar major axis are flat, while minor axis
profiles exhibit steeper slopes. This observational effect of radial mixing is known from main bars, and suggests that
inner bars significantly affect the distribution of stars in the nuclear discs. This effect is consistent with inner
bars being built by stars on strongly elongated orbits, analogous to the $x_1$ orbits in main bars. 

\textbf{(iii)}
The ends of the inner bars exhibit younger mean ages, as compared to the rest of the inner bars. This effect is known
from main bars as kinematic fractionation or orbital age separation and suggests that the youngest stars, i.e.\ those
with the lowest radial velocity dispersion, occupy the most elongated $x_1$ orbits. This effect might as well explain
the formation of $\sigma$-hollows at the ends of the inner bars. 

\textbf{(iv)}
We speculate that the observed young ends of bars and inner bars could be one possible mechanism to create the
morphological feature of ansae, in particular as younger stellar populations have a higher surface brightness at a given
stellar mass surface density. 

The above results, in particular orbital mixing and orbital age separation, together with results from
\citet{deLorenzoCaceres2019} and \citet{mendezAbreu2019} reinforce the idea that barred galaxies can be rather like
Babushka dolls, in which we have a ``galaxy within a galaxy'': inner bars appear to be dynamically very similar to
the main bars of disc galaxies. The main difference between these galaxy components is the spatial scale on which they
form and evolve. 

%
%
%


\begin{acknowledgements}
    We thank the referee for a prompt and constructive report.
    Based on observations collected at the European Southern Observatory under programmes 097.B-0640(A) and
    0103.B-0373(A). The datacubes used in this study are available through the ESO Archive Facility at
    \url{http://archive.eso.org}. 
    TK was supported by the Basic Science Research Program through the National Research Foundation of Korea (NRF)
    funded by the Ministry of Education (No. 2019R1A6A3A01092024). PSB acknowledges financial support from the
    coordinated grants PID2019-107427GB-C31, AYA2016-77237-C3-1-P and AYA2016-77237-C3-2-P from the Spanish Ministry of
    Science, Innovation and Universities (MCIU). AdLC, JF-B, IM-N acknowledge support from grants AYA2016-77237-C3-1-P
    and PID2019-107427GB-C32 of the Spanish Ministry of Science, Innovation and Universities (MCIU) and through the IAC
    project TRACES, which is partially supported by the state budget and the regional budget of the Consejería de
    Economía, Industria, Comercio y Conocimiento of the Canary Islands Autonomous Community.  The Science, Technology
    and Facilities Council is acknowledged by JN for support through the Consolidated Grant Cosmology and Astrophysics
    at Portsmouth, ST/S000550/1.
    This research has made use of the SIMBAD database, operated at CDS, Strasbourg, France \citep{wenger2000}; NASA's
    Astrophysics Data System (ADS); Astropy (\url{http://www.astropy.org}), a community-developed core Python package
    for Astronomy \citep{astropy2013,astropy2018}; NumPy \citet{numpy2006}; SciPy \citep{scipy2020}; and Matplotlib
    \citep{matplotlib2007}.
\end{acknowledgements}

%
%
%


\bibliographystyle{aa}
\bibliography{sections/literature}


\end{document}